\documentclass[conference]{IEEEtran}
\IEEEoverridecommandlockouts

\usepackage{tabularx}
\usepackage{amsmath,amsfonts,amssymb,bm}
\usepackage{graphics} 
\usepackage{epsfig} 
\usepackage{times} 
\usepackage{color}
\makeatletter
\let\NAT@parse\undefined
\makeatother
\usepackage[pagebackref]{hyperref}
\usepackage{xcolor}
\hypersetup{
    colorlinks=true,
    linkcolor={red!75!black},
    citecolor={blue!75!black},
    urlcolor={blue!75!black}
}
\usepackage{cite} 
\usepackage{url}
\usepackage{graphicx,epsfig,epstopdf,float} 
\usepackage{algorithm}
\usepackage[noend]{algpseudocode}
\usepackage{framed}
\usepackage{cancel}
\usepackage[normalem]{ulem}
\usepackage{diagbox}
\usepackage{pifont} 
\usepackage{listings, multicol}
\usepackage{textcomp} 
\usepackage{siunitx} 
\let\labelindent\relax
\usepackage{enumitem}
\usepackage{booktabs}  
\usepackage{subcaption}
\usepackage{multirow}

\definecolor{darkgreen}{RGB}{0,127,0}
\definecolor{darkred}{RGB}{200,0,0}
\definecolor{greentxt}{RGB}{106,168,79}
\definecolor{bluetxt}{RGB}{74,134,232}
\definecolor{redtxt}{RGB}{152,0,0}

\title{\LARGE \bf Determining Sphere Radius through Pairwise Distances}

\author{Boris Sukhovilov  \\
boris.sukhovilov@gmail.com 
}

\begin{document}

\maketitle

\begin{abstract}
We propose a novel method for determining the radius of a spherical surface based on the distances measured between points on this surface. We consider the most general case of determining the radius when the distances are measured with errors and the sphere has random deviations from its ideal shape. For the solution, we used the minimally necessary four points and an arbitrary N number of points. We provide a new closed-form solution for the radius of the sphere through the matrix of pairwise distances. We also determine the standard deviation of the radius estimate caused by measurement errors and deviations of the sphere from its ideal shape. We found optimal configurations of points on the sphere that provide the minimum standard deviation of the radius estimate. This paper describes our solution and provides all the mathematical derivations. We share the implementation of our method as open source code at \url{https://github.com/boris-sukhovilov/Sphere_Radius}.

\end{abstract}


\section{INTRODUCTION}

The sphere radius can be determined from a set of known points on this sphere. We can place \(N\) points on the spherical surface such that not all of them lie on the same non-diametral plane and measure all pairwise distances between the points. A non-diametral plane is a plane that does not intersect the center of the sphere. Our task is to determine the radius of the spherical surface based on the measured pairwise distances between these points.

There are known formulas for calculating the radius of a sphere based on the exact distances between four points. The four points form a tetrahedron and represent the minimally necessary number to solve the problem of determining the radius of the sphere circumscribed around the tetrahedron~\cite{problems1921}.

In practice, there are both errors in distance measurements and deviations of the sphere surface from its ideal shape. The presence of errors and deviations leads to the dependence of the accuracy of the radius estimate on the number of points and their arrangement on the sphere. Thus, four points may not be sufficient to estimate the radius with the required accuracy. Therefore, in the general case, we propose to use a number of points \(N > 4\).

We derived a new formula for calculating the radius through the matrix of pairwise distances between \(N\) points. We determined the standard deviation of the radius estimate. We found optimal configurations of \(N \geq 4\) points on the sphere that provide the minimum standard deviation of the radius estimate.

The structure of this work is as follows. In Section~2, we conduct a literature review. In Section~3, we derive an analytical expression for the radius of the sphere through the pairwise distance matrix between the minimally necessary number of points, which is four points. We then determine the standard deviation of the radius estimate caused by measurement errors. We also find optimal configurations of four points on the sphere that provide the minimum standard deviation of the radius estimate. In Section~4, we develop the radius estimation method for use with \(N \geq 4\) points. We consider possible deviations of the sphere from its shape. We then obtain an analytical expression for the radius of the sphere through the distance matrix between \(N\) points. We determine the standard deviation of the radius estimate caused by measurement errors and deviations of the sphere from its shape. Then we find optimal configurations of \(N\) points on the sphere that provide the minimum standard deviation of the radius estimate. In Section~5, we conduct experiments for our method and evaluate results.

\section{RELATED WORKS}

The calculation of the radius of a sphere circumscribed around a tetrahedron based on exact distances between its vertices is a classic problem in mathematics~\cite{problems1921}. Historically, the first contributions to its solution were made by Carno~\cite{carnot1806}, Brassine~\cite{annales1847}, Crelle~\cite{crelle}, Euler~\cite{novi1758}, Killing, Hovestadt~\cite{killing1913}, and others. Baltzer~\cite{baltzer1861}, Joachimsthal, and Dostor~\cite{joachimsthal1850} used Cayley-Menger determinants for calculating the radius based on exact distances.

However, in practice, the exact distances are not available, and the sphere's surface may have deviations from its ideal shape. This necessitates increasing the number of points to ensure the required accuracy of the radius estimate. We have obtained a new expression for the radius of the sphere through the distance matrix between pairs of \(N \geq 4\) points. This allowed us to study the accuracy of the radius estimate depending on the number and configuration of points on the sphere. We begin our method's examination with the minimally necessary number of points, which is 4.

\section{\textbf{METHOD 1}}
\label{sec:method}

\subsection{\textbf{Estimation of Sphere Radius (N = 4)}}

Let us place N points on the surface of the sphere so that not all of them lie in the same non-diametral plane. We express the radius of the sphere through the measured pairwise distances between the points.

Let the measurement error of the pairwise distances between the specified points and the deviations of the sphere from its shape be so small that it can be neglected. We align the center of the sphere with radius \( R \) with the origin of the coordinate system \( XYZO \).

The distance \( S_{ij} \) between point \( i \) and point \( j \) located on the spherical surface with coordinates \( x_i, y_i, z_i \) and \( x_j, y_j, z_j \) is equal to:

\[ S_{ij} = \sqrt{(x_i - x_j)^2 + (y_i - y_j)^2 + (z_i - z_j)^2}. \tag{1} \]

\noindent Given that \( x_i^2 + y_i^2 + z_i^2 = x_j^2 + y_j^2 + z_j^2 = R^2 \), we rewrite (1) as:

\[ x_i x_j + y_i y_j + z_i z_j = R^2 - \frac{1}{2} S_{ij}^2. \tag{2} \]

\noindent Considering all possible pairwise distances between \( N \) points, we write (2) in matrix form:

\[ \alpha^T \alpha = R_N^2 - S_N^2. \tag{3} \]

\noindent where

\[ \alpha = \begin{bmatrix}
x_1 & x_2 & \cdots & x_N \\
y_1 & y_2 & \cdots & y_N \\
z_1 & z_2 & \cdots & z_N \\
\end{bmatrix} \] – a matrix of size \( 3 \times N \);

\[ R_N^2 = \begin{bmatrix} R^2 & R^2 & \cdots & R^2 \\ R^2 & R^2 & \cdots & R^2 \\ \cdots & \cdots & \cdots & \cdots \\ R^2 & R^2 & \cdots & R^2 \end{bmatrix} \] – a matrix of size \( N \times N \);

\[ S_N^2 = \frac{1}{2} \begin{bmatrix} 0 & S_{12}^2 & S_{13}^2 & \cdots & S_{1N}^2 \\ S_{21}^2 & 0 & S_{23}^2 & \cdots & S_{2N}^2 \\ \cdots & \cdots & \cdots & \cdots &\cdots \\ S_{N1}^2 & S_{N2}^2 & S_{N3}^2 &\cdots & 0 \end{bmatrix} \] – a symmetric matrix of half-squares of distances, size \( N \times N \).

Determine the rank of the matrix \( \alpha^T \alpha \) using Sylvester's inequalities:

\begin{multline*}
\text{rank}(\alpha^T) + \text{rank}(\alpha) - 3 \leq \text{rank}(\alpha^T \alpha) \leq \min(\text{rank}(\alpha^T), \\ \text{rank}(\alpha)). \tag{4}
\end{multline*}

From (4) it follows that \( \text{rank}(\alpha^T \alpha) = 3 \) and all minors of the matrix \( \alpha^T \alpha \) above the third order are zero. We use this property of the matrix \( \alpha^T \alpha \) to find the radius \( R \). Initially, we assume that the first four points out of \( N \) do not lie in one plane.

Let us write the fourth-order minor, formed by the first four columns and rows of the matrix 
\( \alpha^T \alpha \), and equate it to zero:  

\begin{multline*}
\text{det} \begin{bmatrix} 
R^2 & R^2 - \frac{1}{2} S_{12}^2 & R^2 - \frac{1}{2} S_{13}^2 & R^2 - \frac{1}{2} S_{14}^2 \\ 
R^2 - \frac{1}{2} S_{21}^2 & R^2 & R^2 - \frac{1}{2} S_{23}^2 & R^2 - \frac{1}{2} S_{24}^2 \\ 
R^2 - \frac{1}{2} S_{31}^2 & R^2 - \frac{1}{2} S_{32}^2 & R^2 & R^2 - \frac{1}{2} S_{34}^2 \\ 
R^2 - \frac{1}{2} S_{41}^2 & R^2 - \frac{1}{2} S_{42}^2 & R^2 - \frac{1}{2} S_{43}^2 & R^2 
\end{bmatrix} \\ = 0, \tag{5}
\end{multline*}

\noindent where \( \textit{}{det} \) is the determinant of the matrix. 

Using three properties of determinants:
\begin{itemize}
    \item If all elements of a row (column) of the determinant are sums, the determinant can be represented as the sum of the corresponding determinants.
    \item If two rows (columns) of the determinant are proportional to each other, the determinant is zero.
    \item Multiplying all elements of a row (column) of the determinant by some number is equivalent to multiplying the determinant by this number.
\end{itemize}

Transform (5) to this form: 

\begin{multline*}
2 R^2
\left(
\begin{array}{c}
\vspace{0.2cm}
\text{det} \begin{bmatrix} 
1 & S_{12}^2 & S_{13}^2 & S_{14}^2 \\ 
1 & 0 & S_{23}^2 & S_{24}^2 \\ 
1 & S_{32}^2 & 0 & S_{34}^2 \\ 
1 & S_{42}^2 & S_{43}^2 & 0 
\end{bmatrix}\\ \vspace{0.2cm}
+ \text{det} \begin{bmatrix} 
0 & 1 & S_{13}^2 & S_{14}^2 \\ 
S_{21}^2 & 1 & S_{23}^2 & S_{24}^2 \\ 
S_{31}^2 & 1 & 0 & S_{34}^2 \\ 
S_{41}^2 & 1 & S_{43}^2 & 0 
\end{bmatrix} \\ \vspace{0.2cm}
+ \text{det} \begin{bmatrix} 
0 & S_{12}^2 & 1 & S_{14}^2 \\ 
S_{21}^2 & 0 & 1 & S_{24}^2 \\ 
S_{31}^2 & S_{32}^2 & 1 & S_{34}^2 \\ 
S_{41}^2 & S_{42}^2 & 1 & 0 
\end{bmatrix}\\
+ \text{det} \begin{bmatrix} 
0 & S_{12}^2 & S_{13}^2 & 1 \\ 
S_{21}^2 & 0 & S_{23}^2 & 1 \\ 
S_{31}^2 & S_{32}^2 & 0 & 1 \\ 
S_{41}^2 & S_{42}^2 & S_{43}^2 & 1 
\end{bmatrix}
\end{array}
\right)\\ \vspace{0.2cm}
=\text{det} \begin{bmatrix} 
0 & S_{12}^2 & S_{13}^2 & S_{14}^2 \\ 
S_{21}^2 & 0 & S_{23}^2 & S_{24}^2 \\ 
S_{31}^2 & S_{32}^2 & 0 & S_{34}^2 \\ 
S_{41}^2 & S_{42}^2 & S_{43}^2 & 0 
\end{bmatrix}. \tag{6}
\end{multline*}

Consider the determinant on the right-hand side of equation (6). It equals zero only when the four points lie in one plane. In Section 5, we will investigate this case and show that our method works when the points are in the diametral plane or near it. For now, we assume that the matrix determinant has full rank. We divide the left and right-hand sides of equation (6) by the non-zero determinant on the right-hand side (6). Using Cramer's rule, we find that the reciprocal of \( R^2 \) is the sum of solutions \( x_1, x_2, x_3, x_4 \) of the linear equation system of the form: 

\[ 
\frac{1}{2} \begin{bmatrix} 
0 & S_{12}^2 & S_{13}^2 & S_{14}^2 \\ 
S_{21}^2 & 0 & S_{23}^2 & S_{24}^2 \\ 
S_{31}^2 & S_{32}^2 & 0 & S_{34}^2 \\ 
S_{41}^2 & S_{42}^2 & S_{43}^2 & 0 
\end{bmatrix} 
\begin{bmatrix} 
x_1 \\ 
x_2 \\ 
x_3 \\ 
x_4 
\end{bmatrix} = 
\begin{bmatrix} 
1 \\ 
1 \\ 
1 \\ 
1 
\end{bmatrix}. \tag{7}
\]

Introducing the following notations, 

\[ 
\bar{b} = \begin{bmatrix} 
1 \\ 
1 \\ 
1 \\ 
1 
\end{bmatrix}; \quad 
S_{4}^2 = \frac{1}{2} \begin{bmatrix} 
0 & S_{12}^2 & S_{13}^2 & S_{14}^2 \\ 
S_{21}^2 & 0 & S_{23}^2 & S_{24}^2 \\ 
S_{31}^2 & S_{32}^2 & 0 & S_{34}^2 \\ 
S_{41}^2 & S_{42}^2 & S_{43}^2 & 0 
\end{bmatrix}, \tag{8}
\]

\noindent we obtain the formula for calculating the radius of the sphere \( R \):

\[ 
R = \frac{1}{\sqrt{\bar{b}^T (S_{4}^2)^{-1} \bar{b}}}, \tag{9}
\]

\noindent where \( (S_{4}^2)^{-1} \) is the matrix reverse of \( S_{4}^2 \).

\subsection{\textbf{Error Analysis in Estimating Sphere Radius (N = 4)}}

When measuring pairwise distances with errors having zero mean and variance \(\sigma_s^2\), the estimate \(R^*\) of the sphere radius will be:

\[
R^* = \frac{1}{\sqrt{\bar{b}^T (C_4^2)^{-1} \bar{b}}}, \tag{10}
\] where 

\[
C_4^2 = \frac{1}{2} \begin{bmatrix}
0 & C_{12}^2 & C_{13}^2 & C_{14}^2 \\
C_{21}^2 & 0 & C_{23}^2 & C_{24}^2 \\
C_{31}^2 & C_{32}^2 & 0 & C_{34}^2 \\
C_{41}^2 & C_{42}^2 & C_{43}^2 & 0
\end{bmatrix}
\]

\noindent is the matrix of half-squares of pairwise distances measured with errors.

The matrices of true pairwise distances \(S_4\) and measured pairwise distances \(C_4\) are related by: \(C_4 = S_4 + d_4\), where \(d_4\) is the matrix of measurement errors of pairwise distances. The matrices of half-squares of distances \(S_4^2\) and \(C_4^2\) are related by \(C_4^2 = S_4^2 + E_4\), where \(E_4\) is the matrix of measurement errors of the elements of the matrix \(C_4^2\). The matrix \(E_4\) to the second order of smallness is:
\[
E_4 = d_4.*C_4, \tag{11}
\] where .* -  is the element-wise multiplication operation of matrices (Hadamard product of matrices \(d_4\) and \(C_4\)).

Let \(\bar{x}\) be the solution vector of the system:

\[
S_4^2 \bar{x} = \bar{b}, \tag{12}
\]
and \(\bar{x}^* = \bar{x} + \Delta \bar{x}\) be the solution vector of the system:

\[
C_4^2 \bar{x}^* = \bar{b}, \tag{13}
\] where \(\Delta \bar{x}\) is the error vector of the solution due to measurement errors of pairwise distances.  

Subtract (12) from (13):

\begin{multline*}
C_4^2 \bar{x}^* - (C_4^2 - E_{4})(\bar{x}^* - \Delta \bar{x}) = E_{4} \bar{x}^* + C_4^2 \Delta \bar{x} - E_{4} \Delta \bar{x} \\ = 0, 
\tag{14}
\end{multline*}

\noindent Neglecting the vector \(E_{4} \Delta \bar{x}\) with elements of the second-order smallness, we obtain the error vector of the solution of system (13) as:

\[
\Delta \bar{x} = - (C_4^2)^{-1} E_{4} \bar{x}^*. \tag{15}
\]

\noindent Multiplying both sides of equality (15) from the left by \(\bar{b}^T\), we get the error value of \(\frac{1}{R^{*2}} \):

\[
\Delta (\frac{1}{R^{*2}}) = \bar{b}^T \Delta \bar{x} = - \bar{b}^T (C_4^2)^{-1} E_{4} \bar{x}^* = - \bar{x}^{*T} E_{4} \bar{x}^*. \tag{16}
\]

\noindent Let the variance of \(\frac{1}{R^{*2}} \) be denoted by \(D\). Then \(D\) is defined as:

\[
D = \mathbb{E}[\Delta (\frac{1}{R^{*2}})^2] = \mathbb{E}[(\bar{x}^{*T} E_{4} \bar{x}^*)^2], \tag{17}
\] where \(\mathbb{E}[.]\) is the expectation operator. Substituting matrix \(E_{4}\) from (11) into (17), we get the variance:

\[
D = 2 \sigma_{S}^2 Y^{*T} C_{4}^2 Y^*, \tag{18}
\] where vector \(Y^* \) consists of squares of the components of vector \(\bar{x}^*\).

Let us find the standard deviation of the estimate of \( R^* \).
Consider the function inverse to the square of the radius \( f = \frac{1}{R^2} \).
Let us express \( R \) as a function of \( f \): \( R = f^{-\frac{1}{2}} \) and find the linear increment \( \Delta R \) at the point of the computed value \( f^* = \frac{1}{R^{*2}} \)

\[ \Delta R = -\frac{1}{2} f^{* - \frac{3}{2}} \Delta f = -\frac{1}{2} (R^*)^3 \Delta f. \tag{19}\]

\noindent From (19) we get the standard deviation of the estimate of \( R^* \):

\[ \sigma_R = \sqrt{\mathbb{E}[\Delta R^2]} = \frac{1}{2} (R^*)^3 \sqrt{D}.\tag{20} \]

\subsection{\textbf{Determining Optimal Point Configurations (N = 4)}}

As follows from (18) and (13), the variance \( D \) depends on the variance \( \sigma^2_s \) and the matrix \( C^2_4 \). The matrix \( C^2_4 \) defines the geometry of the arrangement of four points on the sphere, forming a tetrahedron. The geometry of the tetrahedron is determined by its edges – pairwise distances \( S_{12}, S_{13}, S_{14}, S_{23}, S_{24}, S_{34} \). Therefore, reducing the variance \( D \), and thus the standard deviation of the sphere radius estimate \( \sigma_R \), can be achieved not only by increasing the measurement accuracy but also by selecting optimal pairwise distances.

To determine the pairwise distances that provide the minimum variance \( D \), we minimize the functional

\[
D = 2\sigma^2_s Y^T S^2_4 Y \xrightarrow[\bar{x}, S^2_4]{} \min, \tag{21}
\]

subject to the constraints:

\[
S_4^2 \bar{x} = \bar{b}; \tag{22}
\]

\[
\bar{b}^T \bar{x} = \frac{1}{R^2}. \tag{23}
\]

By solving the problem using the method of Lagrange multipliers, we establish that the variance $D$ has a global minimum 
\[
D_0 = \frac{1}{R^6} \cdot \frac{\sigma_s^2} {8}, \tag{24}
\]

\noindent when the components of the vector \( \bar{x} \) are equal, and the pairwise distance matrix \( S_{4} \) has the following form

\[
S_{4} = \begin{bmatrix}
0 & a & b & \ell \\
a & 0 & \ell & b \\
b & \ell & 0 & a \\
\ell & b & a & 0
\end{bmatrix},\tag{25}
\]

\noindent where \( a = S_{12} = S_{34} \); \( b = S_{13} = S_{24} \); \( \ell = S_{23} = S_{14} \).

From (25), it follows that the optimal configurations of 4 points on the sphere form tetrahedra with pairwise equal opposite edges and, in general, unequal non-opposite edges. The triangles forming the faces of such tetrahedra are equal. There are an endless number of equilateral tetrahedra that can be inscribed in a given sphere (with all vertices lying on this sphere), resulting in an infinite number of optimal configurations. Let us denote the equal components of the vector \( \bar{x} \) by the variable \( x_0 \). Substituting the vector \( \bar{x} \), composed of equal components, into equation (23), we obtain the value $x_0 = \frac{1}{4R^2}$. Substituting the value \( x_0 \) into (22)

\[
S^2_4 \bar{b} = 4R^2 \bar{b} \tag{26}
\]
\noindent From (26), it follows that the vector \( \bar{b} \) (for optimal configurations of points on the sphere) is an eigenvector of the matrix \( S^2_4 \). It corresponds to its largest eigenvalue \( NR^2 \) for \( N = 4 \). Obviously, the vector \( \bar{b} \) is also an eigenvector of the matrix \( (S^2_4)^{-1} \) and corresponds to its smallest eigenvalue in absolute value \( \frac{1}{4R^2} \).

Multiplying both sides of equation (26) on the left by the vector \( \bar{b}^T \). Replacing the true pairwise distance matrix \( S^2_4 \) with the measured matrix \( C^2_4 \), we obtain the formula for estimating the sphere radius for optimal configurations of 4 points on the sphere:
\[
R^*_o = \frac{\sqrt{\bar{b}^T C^2_4 \bar{b}}}{4} \tag{27}
\]

\noindent In (27), unlike in (10), there is no matrix inversion operation. Therefore, the calculation of the sphere radius using (27) is stable even when the matrix \( S^2_4 \) has an incomplete rank, and the matrix \( C^2_4 \) is thus ill-conditioned. Critical configurations, when the matrix \( S^2_4 \) has an incomplete rank, will be discussed in Section 5.

Another general stable method for calculating the sphere radius is our method, which we introduce in the next section. We call it "general stable" because it works and is stable for \( N \geq 4 \) points. The method takes into account random deviations of the sphere from its shape.

\section{\textbf{METHOD 2}}
\subsection{\textbf{Estimation of Sphere Radius Considering Shape Deviation $N \geq 4$}}

Let us generalize our method for estimating the radius of the sphere for \( N \geq 4 \) points fixed on the sphere and consider possible random deviations of the sphere from its shape. We define deviations as the difference between the radius \( R \) and the distance from the center of the sphere to a placed point. We assume that the deviations are random, independent of the errors in measuring distances between points, have zero mean, and variance \( \sigma_m^2 \). The center of the approximating spherical surface coincides with the origin of the XYZO coordinate system.

Write the equations (12), (13) for \( N \) points as:

\[
S^2_N \bar{x} = \bar{b}, \tag{28}
\]
\[
C^2_N \bar{x}^* = \bar{b}, \tag{29}
\]

\noindent where $
S^2_N = \frac{1}{2} \begin{bmatrix}
0 & S_{12}^2 & \dots & S_{1N}^2 \\
S_{21}^2 & 0 & \dots & S_{2N}^2 \\
\dots & \dots & \dots & \dots \\
S_{N1}^2 & S_{N2}^2 & \dots & 0
\end{bmatrix}$  is the matrix of half-squares of true distances between \( N \) points on the sphere,
\( \bar{x} \) is an \( N \)-dimensional solution vector of system (28),

$
C^2_N = \frac{1}{2} \begin{bmatrix}
0 & C_{12}^2 & \dots & C_{1N}^2 \\
C_{21}^2 & 0 & \dots & C_{2N}^2 \\
\dots & \dots & \dots & \dots \\
C_{N1}^2 & C_{N2}^2 & \dots & 0
\end{bmatrix}
$  is the matrix of half-squares of measured distances with errors between \( N \) points on the sphere with deviations,
\( \bar{x}^* \) is the \( N \)-dimensional solution vector of system (29),
\( \bar{b} \) is a vector consisting of \( N \) ones.

The rank of the matrix \( S^2_N \) is 4. Therefore, the measured matrix \( C^2_N \) also has a rank 4. We use the approach we employed for 4 points to estimate the radius. We seek the quantity, inverse to the square of the sphere radius, as the sum of the elements of the vector \( \bar{x}^* \), which is the solution to the system of linear equations (29). For \( N > 4 \), the system (29) has an incomplete rank. Therefore, to compute the vector \( \bar{x}^* \), we use the pseudo-inversion operation of the matrix \( C^2_N \), equating \( N - 4 \) of its smallest eigenvalues in absolute value to zero. These eigenvalues are due to errors in the original data. Using pseudo-inversion, we filter these errors. As a result, we obtain the formula for calculating the sphere radius through the measured matrix of half-squares of pairwise distances \( C^2_N \)

\[
R^* = \frac{1}{\sqrt{\bar{b}^T (C_N^2)^+ \bar{b}}}, \tag{30}
\]
\noindent where \((C^2_N)^+\) is the pseudo-inverse matrix for \( C^2_N \). In our programs, we use singular value decomposition to compute \((C^2_N)^+\).

In section 5, we explore configurations of points on the sphere where the rank of the matrix $S^2_N$ is less than 4 and demonstrate that our method solves the problem even in these cases.

Let us assess the accuracy of the obtained solution. To do this, we calculate the standard deviation of the sphere radius estimate $R^*$. Suppose that

\[
C^2_N = S^2_N + E + \Phi, \tag{31}
\]

\noindent where $E$ is the error matrix of measuring half-squares of pairwise distances, $\Phi$ is the error matrix caused by deviations of the sphere from its shape.

The elements of the matrix $E$ in the linear approximation are equal to

\[
E = d_N .* C_N, \tag{32}
\]

\noindent where $C_N$ is the matrix of measured pairwise distances, $.*$ is the element-wise multiplication operation of matrices, $d_N$ is the error matrix of measuring pairwise distances. The elements of the matrix $d_N$ are independent, centered, and have variance $\sigma^2_s$.

We compute the matrix $\Phi$. The square of the distance $S^2_{ij}$ between points $i$ and $j$ on the sphere with radius vectors of coordinates $\bar{r}_i = [x_i, y_i, z_i]$, $\bar{r}_j = [x_j, y_j, z_j]$ is equal to

\[
S^2_{ij} = r^2_i + r^2_j - 2 r_i r_j \cos \beta_{ij}, \tag{33}
\]

\noindent where $r_i$ and $r_j$ are distances from the origin of the $XYZO$ coordinate system to points $i$ and $j$; $\beta_{ij}$ is the angle between the radius vectors $\bar{r}_i$ and $\bar{r}_j$. The element of the matrix $\Phi_{ij}$ is equal to half the error $\Delta S^2_{ij}$ caused by random deviations $\Delta r_i = r_i - R$, $\Delta r_j = r_j - R$ of the sphere from its shape, and in the linear approximation is

\begin{multline*}
\Phi_{ij} = \frac{1}{2} \Delta S^2_{ij} = R (\Delta r_i + \Delta r_j) (1 - \cos \beta_{ij}) = \\2 R \left( \sin \frac{\beta_{ij}}{2} \right)^2 (\Delta r_i + \Delta r_j) = \frac{S^2_{ij}}{2R} (\Delta r_i + \Delta r_j). \tag{34}
\end{multline*}

Subtracting (28) from (29) and performing transformations similar to (14), (15)

\[
\bar{x}^* - \bar{x} = - (C^2_N)^+ (E + \Phi) \bar{x}^*, \tag{35}
\]

\noindent we obtain the error in the estimate of the quantity $\frac{1}{R^{*2}}$ by multiplying both sides of equation (35) on the left by $\bar{b}^T$:

\begin{align}
& \Delta(\frac{1}{R^{*2}}) = \bar{b}^T\bar{x}^* - \bar{b}^T\bar{x} = -\bar{b}^T(C_{N}^2)^+(E+\Phi)\bar{x}^* \nonumber\\&= 
\bar{x}^{*T}(E+\Phi)\bar{x}^* =
\bar{x}^{*T} E \bar{x}^* + \bar{x}^{*T} \Phi \bar{x}^* = \bar{x}^{*T}E\bar{x}^* \nonumber\\&
+ \frac{2}{R^*}
\left(
\begin{array}{c}
\vspace{0.2cm}
\bar{x}^{*T}
\begin{bmatrix}
\Delta r_1 & 0 & \cdots & 0 \\
0 & \Delta r_2 & \cdots & 0 \\
\dots & \dots & \cdots & \dots \\
0 & 0 & \cdots & \Delta r_N 
\end{bmatrix}\nonumber\\ 
{\frac{1}{2}} 
\begin{bmatrix}
0 & C_{12}^2 & \cdots & C_{1N}^2 \\
C_{21}^2 & 0 & \cdots & C_{2N}^2 \\
\dots & \dots & \cdots & \dots \\
C_{N1}^2 & C_{N2}^2 & \cdots & 0 
\end{bmatrix} 
\bar{x}^*
\end{array}
\vspace{0.2cm}
\right)\nonumber\\ 
&= \bar{x}^{*T}E\bar{x}^* + \frac{2}{R^*}
\bar{x}^{*T}
\begin{bmatrix}
\Delta r_1 & 0 & \cdots & 0\\
0 & \Delta r_2 & \cdots & 0\\
\dots & \dots & \cdots & \dots\\
0 & 0 & \cdots & \Delta r_N
\end{bmatrix}\bar{b}\nonumber\\
&= \bar{x}^{*T}E\bar{x}^* + \frac{2}{R^*}
\Delta r^T\bar{x}^*, \tag{36}
\end{align}

\noindent where the vector of deviations of the sphere from the shape $\Delta \mathbf{r}^T = [\Delta r_1 \Delta r_2 \ldots \Delta r_N]$, and $C^2_N \bar{x}^* = \overline{\bar{b}}$.

We raise $\Delta\left(\frac{1}{R^*2}\right)$ to the square. We perform the expectation operation, considering the independence of the measurement errors of distances and deviations of the sphere from the shape. As a result, we obtain the variance of the quantity $\frac{1}{R^{*2}}$:

\[
D_N = 2\sigma_s^2Y^{*T}C_N^2Y^* + \frac{4\sigma_m^2}{R^{*2}}\bar{b}^TY^*, \tag{37}
\]

\noindent where the vector $Y^*$ is composed of the squares of the components of the vector $\bar{x}^*$, $\sigma_s^2$ is the variance of the measurement errors of pairwise distances, and $\sigma_m^2$ is the variance of the sphere's deviations from its shape.

As in deriving (20), the standard deviation of the sphere radius estimate $R^*$ is determined by substituting the value of the variance $D_N$ into the formula 
$
\sigma_R = \frac{1}{2}(R^*)^3\sqrt{D_N}.
$

\subsection{\textbf{Determining Optimal Configurations of Points on a Sphere with Shape Deviations $(N \ge 4)$}}

To determine the optimal configurations of $N$ points on a sphere with shape deviations, we minimize the functional

\[
D_N = 2\sigma_s^2 Y^T S_N^2 Y + \frac{4\sigma_m^2}{R^2} \bar{b}^T Y^T \xrightarrow[\bar{x}, S_N^2]{} \min \tag{38}
\]

\noindent subject to the constraints:

\[
S_N^2 \bar{x} = \bar{b}, \tag{39}
\]

\[
\bar{b}^T \bar{x} = \frac{1}{R^2}. \tag{40}
\]

\noindent By solving the problem, we established that the global minimum

\[
D_{N0} = \frac{1}{R^6}(\frac{2\sigma_s^2}{N^2} + \frac{4\sigma_m^2}{N}), \tag{41}
\]

\noindent occurs when the components of the vector $\bar{x}$ are equal.

Let us denote, as before, the equal components of the vector $\bar{x}$ by the variable $x_0$. Substituting into equation (40) the vector $\bar{x}$, composed of equal components, we obtain the value $x_0 = \frac{1}{N R^2}$. Substituting the value $x_0$ into (39):

\[
S_N^2 \bar{b} = N R^2 \bar{b}. \tag{42}
\]

From (42), it follows that the vector $\bar{b}$ (for optimal configurations of points on the sphere) is an eigenvector of the matrix $S_N^2$. It corresponds to its largest eigenvalue $N R^2$.

In equation (42), the left side represents the sum of the half-squares of the distances from each point to all others, and the right side is a constant vector with components equal to $N R^2$. Therefore, for optimal configurations of $N$ points, the sum of the squares of the distances from each point to all others is constant. Later, in section 5, we consider the geometry of such shapes.

Multiplying both sides of equation (42) on the left by the vector $\bar{b}^T$ and replacing the matrix $S_N^2$ with its measured value, we obtain another formula for estimating the radius for optimal configurations of points on the sphere:

\[
R_0^* = \frac{\sqrt{\bar{b}^T C_N^2 \bar{b}}}{N}. \tag{43}
\]

The standard deviation of the radius estimate $R^*$ for optimal configurations of points on the sphere is:

\[
\sigma_R = \frac{1}{2}(R^*)^3\sqrt{D_{N0}} = \sqrt{\frac{\sigma_s^2}{2N^2} + \frac{\sigma_m^2}{N}}. \tag{44}
\]

From (44), it follows that for optimal configurations of points on the sphere, the radius estimate is asymptotically efficient with respect to the number of points between which pairwise distances are measured.

\section{Analysis of the Method and Experimental Studies}

Let us analyze the conditions under which our method allows estimating the sphere radius when the rank of the matrix $S_N^2$ becomes less than four. Consider the optimal configuration of 4 points on the sphere, for which, as follows from (25), the matrix of the half-squares of distances is

\[
S_4^2 = \frac{1}{2}\begin{bmatrix}
0 & a^2 & b^2 & \ell^2 \\
a^2 & 0 & \ell^2 & b^2 \\
b^2 & \ell^2 & 0 & a^2 \\
\ell^2 & b^2 & a^2 & 0 
\end{bmatrix}. \tag{45}
\]

\noindent Calculate the eigenvalues of the matrix $S_4^2$:

\begin{align}
&\lambda_1 = \frac{1}{2}(-a^2 - b^2 + \ell^2) \nonumber\\
&\lambda_2 = \frac{1}{2}(-a^2 + b^2 - \ell^2) \nonumber\\
&\lambda_3 = \frac{1}{2}(a^2 - b^2 - \ell^2) \nonumber\\
&\lambda_4 = \frac{1}{2}(a^2 + b^2 + \ell^2) \tag{46}
\end{align}

\noindent Using (27) we calculate the square of the sphere radius:

\[
R^2 = \frac{1}{8}(a^2 + b^2 + \ell^2). \tag{47}
\]

\noindent The matrix $S_4^2$ has an incomplete rank, less than four, if one of the first three eigenvalues equals zero. Consider this condition for the eigenvalue $\lambda_1$. Setting (46) to zero and considering (47), we get:

\[
\lambda_1 = \frac{1}{2}(-a^2 - b^2 + \ell^2) = 0 \rightarrow \ell^2 = a^2 + b^2 \rightarrow \ell = 2R. \tag{48}
\]

\noindent Formula (48) describes a right triangle with the hypotenuse equal to the diameter of the sphere. This occurs when all 4 points lie in a diametral plane, and the tetrahedron degenerates into a rectangle with the diagonal equal to the diameter of the sphere. In this case, the rank of the matrix $S_4^2$ is 3. When $b = \ell$, $a \rightarrow 0$ the rank of $S_4^2$ tends to 2. The trace of the matrix $S_4^2$ equals 0. The sum of the eigenvalues of $S_4^2$ is equal to its trace. Therefore, the rank of $S_4^2$ cannot be 1. Thus, the rank of the matrix $S_4^2$ can take values ranging from 4 to 2.

Note that when all points are located in a non-diametral plane, we can only estimate the radius of the circle on which the points are located, but not the radius of the sphere. When estimating the sphere radius, we should not allow such a point configuration.

When points are located in the diametral plane or near it, formula (30) allows calculating the radius of the diametral circle and, accordingly, the sphere radius, as it uses the pseudo-inversion of the matrix $C_N^2$.

When performing pseudo-inversion, it is important to correctly estimate the rank of the matrix $C_N^2$. The complexity of the task is due to the presence of errors in the elements of the matrix $C_N^2$. We estimate the rank of $C_N^2$ by analyzing the reliability of its four largest eigenvalues in absolute value. The initial data for analysis is the variance of the measurement errors of pairwise distances. In the case of sphere deviations from shape, the variance of sphere deviations from shape and an initial approximation of the sphere radius $R_0$ are required. An example of choosing $R_0$ will be considered below.

The eigenvalues $\lambda_i^*$, $\lambda_i$, (i = 1, 2, 3, 4) of the matrices $C_N^2$, $S_N^2$ are related by:

\[
\lambda_i^* = \lambda_i + \Delta \lambda_i
\]

\noindent where increments $\Delta \lambda_i$, (i = 1, 2, 3, 4) are caused by the total perturbation matrix $(E + \Phi)$, according to (31). Given the symmetry of the matrices $C_N^2$, $S_N^2$, in the linear approximation, the value of $\Delta \lambda_i$ can be estimated as \cite{faddeev1961}, paragraph 41:

\[
\Delta \lambda_i = \psi_i^T [(E + \Phi)] \psi_i, \tag{49}
\]

\noindent where $\psi_i$ is the eigenvector of the matrix $C_N^2$, corresponding to the eigenvalue $\lambda_i^*$.

\noindent We perform transformations (49):

\begin{multline}
\Delta \lambda_i = \psi_i^T[(E + \Phi)] \psi_i
= \psi_i^T E \psi_i + \psi_i^T \Phi \psi_i = \psi_i^T E \psi_i + \\
\frac{2}{R_0}
\left(
\begin{array}{c}
\vspace{0.2cm}
\psi_i^T
\begin{bmatrix}
\Delta r_1 & 0 & \cdots & 0 \\
0 & \Delta r_2 & \cdots & 0 \\
\dots & \dots & \cdots & \dots \\
0 & 0 & \cdots & \Delta r_N
\end{bmatrix}\\
\frac{1}{2}
\begin{bmatrix}
0 & C_{12}^2 & \cdots & C_{1N}^2 \\
C_{21}^2 & 0 & \cdots & C_{2N}^2 \\
\cdots & \dots & \cdots & \dots \\
C_{N1}^2 & C_{N2}^2 & \cdots & 0
\end{bmatrix}
\psi_i
\vspace{0.2cm}
\end{array}
\right)\\
= \psi_i^T E \psi_i + \frac{2 \lambda_i^*}{R_0}\ \psi_i^T
\begin{bmatrix}
\Delta r_1 & 0 & \cdots & 0 \\
0 & \Delta r_2 & \cdots & 0 \\
\dots & \dots & \cdots & \dots \\
0 & 0 & \cdots & \Delta r_N
\end{bmatrix}
\psi_i = \\ 
\psi_i^T E \psi_i + \frac{2 \lambda_i^*}{R_0} \Delta r^T \gamma_i, \tag{50}
\end{multline}

\noindent where the vector $\gamma_i = \psi_{i\cdot} * \psi_i$ is composed of the squares of the elements of the vector $\psi_i$, and $C_N^2 \psi_i = \lambda_i^* \psi_i$.

Raise $\Delta \lambda_i$ to the square, perform the expectation operation considering the independence of the measurement errors of distances and deviations of the sphere from shape. As a result, we obtain the variance of the quantity $\Delta \lambda_i$

\[
D_{\Delta \lambda_i} = 2\sigma_s^2 \gamma_i^T C_N^2 \gamma_i + \frac{4\sigma_m^2 \lambda_i^{*2}}{R_0^2} \bar{b}^T (\gamma_{i\cdot} * \gamma_i). \tag{51}
\]

We consider the normal distribution laws of the measurement errors of distances and random deviations of the sphere from shape. Then, in the linear approximation, the distribution law of $\Delta \lambda_i$ will be normal with zero mean and variance $D_{\Delta \lambda_i}$. This allows establishing the following rule for calculating the eigenvalues $\lambda_i^{*-1}$ for the pseudo-inverse matrix $(C_N^2)^+$:

\begin{multline}
\text{If } |\lambda_i^*| > K_p \sqrt{D_{\Delta \lambda_i}} + {Tol}, \text{ then } \lambda_i^{*-1} = \frac{1}{\lambda_i^*},\\
\text{ else } \lambda_i^{*-1} = 0, \tag{52}
\end{multline}

\noindent where $K_p \sqrt{D_{\Delta \lambda_i}}$ is the random component of the reliability threshold $\lambda_i^*$ at the confidence level $p$. In our calculations, we adopted $K_p = 4$ at $p = 99.99\%$. $Tol$ is the component of the reliability threshold associated with machine computation errors.

If the value of the eigenvalue $\lambda_i^*$ is less than the reliability threshold, we consider the calculated value $\lambda_i^*$ unreliable and lower the rank of the matrix $C_N^2$. Accordingly, when calculating the pseudo-inverse matrix $(C_N^2)^+$, we equate its eigenvalue $\lambda_i^{*-1}$ to $0$, corresponding to the eigenvalue $\lambda_i^*$ of the matrix $C_N^2$.

Consider the following example provided by Yu. S. Vasiliev. Four points are placed on a sphere of unit radius, located in two mutually perpendicular diametral planes. Two points are at the north pole and two at the south pole. The points are displaced from the poles by the same polar angle, equal to $0.0988$ radians. The six pairwise distances $S_{12} = 0.19728; S_{13} = 1.9951; S_{14} = 1.9951; S_{23} = 1.9951; S_{24} = 1.9951; S_{34} = 0.19728$ are measured with relative errors: $\delta_{12} = 0.01; \delta_{13} = 0.01; \delta_{14} = 0; \delta_{23} = 0; \delta_{24} = 0; \delta_{34} = 0.01$. Accordingly, the measured distances are: $C_{12} = 0.19925; C_{13} = 2.0151; C_{14} = 1.9951; C_{23} = 1.9951; C_{24} = 1.9951; C_{34} = 0.19925$.

Based on the values of relative errors, we have designated the input parameters of the algorithm for calculating the radius of the sphere. This includes the standard deviation of the measured distances and the deviations of the sphere from the shape: $\sigma_s = 0.01; \sigma_m = 0.01$. The initial approximation of the radius $R_0$, which is used in calculating the contribution of the deviations of the sphere from the shape in the evaluation of $D_{\Delta \lambda_i}$, was set (for verification) with a deviation from the true radius by $50\%$ in both smaller and larger directions. It should be noted that these variations of $R_0$ did not affect the calculation results.

Table I shows the results of the sphere radius calculations, performed by six methods.

\begin{table}[h]
\centering
\begin{tabularx}{0.48\textwidth}{|>{\centering\arraybackslash}X|>{\centering\arraybackslash}X|}
\hline
Method & Radius Estimate \\
\hline
Carnot formula & 0.587799 \\
\hline
Euler, Grelle formulas & 0.587799 \\
\hline
Cayley-Menger determinant formulas & 0.587799 \\
\hline
Our method formula (10) & 0.587799 \\
\hline
Our method formula (27) & 1.00255 \\
\hline
Our method formula (30) & 1.00257 \\
\hline
\end{tabularx}
\caption{Sphere Radius Estimate (R=1)}
\end{table}

The first three methods turned out to be sensitive to errors in the initial data. The reason is that all 4 points are located near one straight line, which is the diameter of the sphere. The volume of the tetrahedron formed by these points is close to 0. This leads to catastrophic errors in calculating the radius of the sphere circumscribed around the tetrahedron. In the fourth method, catastrophic errors in calculations are caused by the inversion of an ill-conditioned matrix $C_N^2$. The fifth method showed good results, as this configuration of 4 points forms a tetrahedron with nearly equal opposing edges. The method is designed for such cases and does not use the inversion of the matrix $C_N^2$ in calculating the radius of the sphere. The sixth, general method, also showed good results since it correctly calculated the rank of the matrix $C_N^2$, which is equal to 2. The calculation results are confirmed by the analysis of formulas (46). In our example, $b \approx \ell$ and the absolute values of the eigenvalues $\lambda_1^* \approx -\frac{1}{2} a^2; \lambda_2^* \approx -\frac{1}{2} a^2$ are less than the calculated overall threshold, guaranteeing their reliability. Accordingly, when calculating the pseudo-inverse of the matrix $(C_N^2)^+$, its eigenvalues $\lambda_1^{*-1}, \lambda_2^{*-1}$ are set to zero. This ensures the correct solution to the problem.

The supplemental program implementing the example is located at \url{https://github.com/boris-sukhovilov/Sphere_Radius} in the file \texttt{example\_for\_suh.m}. The method for computing the radius  is in the file \texttt{calc\_Radius.m}.

Now let us consider the optimal configurations of $N$ points on a sphere and calculate their radius. As noted in paragraph IV.B, for optimal configurations of $N$ points, the sum of the squares of the distances from each point to all others is constant. Five regular convex polyhedra, known as Platonic solids, satisfy this rule. These are the regular tetrahedron, octahedron, hexahedron, icosahedron, and dodecahedron. A program that implements the radius calculation for the sphere described around Platonic solids is in the file \texttt{regular\_polyhedron.m}.

Besides regular convex polyhedra, there are countless irregular figures where the sum of the squares of distances from each vertex to all others is constant. To calculate the coordinates of $N$ vertices of irregular figures inscribed in the sphere, we developed the program \texttt{optimal\_n\_points\_on\_sphere\_fix.m}. The calculation algorithm is based on solving an optimization problem using the Nelder-Mead method. First, we generate an $N$-dimensional random simplex of points on the sphere, divided into two groups. The division is necessary to exclude the generation of simplexes where all vertices are located near the same non-diametral plane. The first group consists of 4 vertices forming a tetrahedron. One vertex of the tetrahedron is fixed at the north pole. From the three vertices of the tetrahedron base, we form a regular triangle. We place it in a random plane perpendicular to the diameter passing through the north and south poles of the sphere. The horizontal position of the plane is randomly varied within the range of half the sphere's radius length up and down from its center. The azimuthal angle of the second vertex is fixed so that it is always located in the $OXZ$ plane. The remaining points of the second group are randomly placed according to a uniform distribution within the ranges of azimuthal and polar angles encompassing the entire sphere. The adopted rules for fixing the first and second vertices of the tetrahedron make the simplex maintain its position in space during optimization. The initial lengths of the simplex edges are random but ensure that the simplex vertices are located on the sphere. The simplex retains degrees of freedom to change the lengths of its edges. As a result of minimizing the variance (37), we obtain the final optimal simplex from the initial simplex. We then model random deviations of the simplex vertices from the sphere's surface and measure the distances between the simplex vertices with errors. We use the matrix of measured pairwise distances to calculate the sphere's radius. The calculation results are presented in Table II. Columns such as Euler, Grelle formulas, and Cayley-Menger determinant formulas are removed from the table since their values coincide with the value of the Carnot formula. The best results are highlighted in bold.

\begin{table}[h]
\centering
\begin{tabularx}{0.48\textwidth}{|>{\centering\arraybackslash}X|>{\centering\arraybackslash}X|>{\centering\arraybackslash}X|>{\centering\arraybackslash}X|>{\centering\arraybackslash}X|>{\centering\arraybackslash}X|}
\hline
\multirow{2}{=}{\centering Number\\of points} & \multicolumn{5}{c|}{Method} \\
\cline{2-6}
 & Our method formula (30), mm & Eberly method, mm & Sumith method, mm & Our method formula (43), mm & Carnot formula, mm \\
\hline
4 & \textbf{0.2038} & 0.2159 & 1.4337 & 0.2227 & 1.4154 \\
\hline
5 & \textbf{0.1613} & 0.1999 & 0.1995 & 0.1614 & – \\
\hline
10 & \textbf{0.08303} & 0.1255 & 0.1255 & \textbf{0.08303} & – \\
\hline
20 & \textbf{0.03795} & 0.06625 & 0.06626 & \textbf{0.03795} & – \\
\hline
50 & 0.01642 & 0.02757 & 0.02757 & \textbf{0.01641} & – \\
\hline
\end{tabularx}
\caption{The Mean Absolute Error of Radius Estimation (R=1000 mm) of the Sphere Optimal Arrangement of Points on the Entire Sphere, $\sigma_s = 1 \, \text{mm}; \, \sigma_m = 1 \, \text{mm}$}
\end{table}

In the calculations, for comparison, we used the methods for estimating the sphere's radius described in the literature ~\cite{eberly2024, sy1506}. These methods use the coordinates of points on the sphere as initial data. Therefore, before applying them, we calculate the coordinates of the points based on the measured pairwise distances. Let's consider these calculations. For this, we represent the points on the sphere as a mechanical structure of material points with unit mass, connected by weightless rigid edges. We place the origin of the coordinate system at the center of gravity and direct the coordinate axes along the inertia axes of the structure. This coordinate system is well-balanced to minimize the errors in calculating the points' coordinates. The program that calculates the points' coordinates based on the matrix of measured pairwise distances in our adopted coordinate system is in the file \texttt{CenterOfGravityCoordFromPairDistance.m}.

As shown by the data in Table II, our methods for estimating the radius using formulas (30) and (43) have demonstrated the best accuracy results.

Next, we conducted statistical tests of our method with random placement of 4, 5, 10, 20, 50, 100 points on the surface of a sphere of different extents. The extent was determined by the ranges of azimuthal and polar angles. The range of the azimuthal angle was set to the maximum possible. The variation of the sphere's extent was ensured by varying the upper value of the polar angle range.

With the random placement of points on the sphere, situations are possible where all the points are located near the same non-diametral plane. In this case, we can only estimate the radius of the circle on which the points are located, but not the sphere's radius. This leads to the inability to reliably estimate the accuracy of the calculated radius. To avoid this, we divide the points of the generated simplex into two groups. The first group consists of 4 vertices forming a tetrahedron. One vertex of the tetrahedron is fixed at the north pole. From the three vertices of the tetrahedron base, we form a regular triangle. We place it in a plane perpendicular to the diameter passing through the north and south poles. The plane below cuts off the surface of the sphere, on which we place the remaining points of the second group. These points are placed randomly on the surface using a uniform distribution within the designated ranges of azimuthal and polar angles. This configuration of points excludes situations where all the points are located near the same non-diametral plane.

The program that implements the statistical tests is located in the file \texttt{SphereRadiusTest.m}. The calculation results are presented in Tables III, IV, V, and VI. Columns such as Euler, Grelle formulas, and Cayley-Menger determinant formulas are removed from the tables since their values coincide with the value of the Carnot formula. The best results are highlighted in bold.

\begin{table}[h]
\centering
\begin{tabularx}{0.48\textwidth}{|>{\centering\arraybackslash}X|>{\centering\arraybackslash}X|>{\centering\arraybackslash}X|>{\centering\arraybackslash}X|>{\centering\arraybackslash}X|}
\hline
\multirow{2}{=}{\centering Number\\of points} & \multicolumn{4}{c|}{Method} \\
\cline{2-5}
 & Our method formula (30), mm & Eberly method, mm & Sumith method, mm & Carnot formula, mm \\
\hline
4 & \textbf{4.4690} & 34.0204 & \textbf{4.4690} & \textbf{4.4690} \\
\hline
5 & \textbf{4.3466} & 43.0431 & 4.3720 & - \\
\hline
10 & \textbf{3.4993} & 70.9088 & 3.5155 & - \\
\hline
20 & \textbf{2.5248} & 90.4502 & 2.5249 & - \\
\hline
50 & \textbf{1.3197} & 106.946 & 1.3488 & - \\
\hline
100 & \textbf{0.9238} & 111.329 & 0.9297 & - \\
\hline
\end{tabularx}
\caption{The Mean Absolute Error of Radius Estimation (R=1000 mm) of the Sphere
Angle Ranges: Polar [0 45°], Azimuthal [0 360°]; $\sigma_s = 1 \, \text{mm}; \, \sigma_m = 1 \, \text{mm}$}
\end{table}

\begin{table}[h]
\centering
\begin{tabularx}{0.48\textwidth}{|>{\centering\arraybackslash}X|>{\centering\arraybackslash}X|>{\centering\arraybackslash}X|>{\centering\arraybackslash}X|>{\centering\arraybackslash}X|}
\hline
\multirow{2}{=}{\centering Number\\of points} & \multicolumn{4}{c|}{Method} \\
\cline{2-5}
 & Our method formula (30), mm & Eberly method, mm & Sumith method, mm & Carnot formula, mm \\
\hline
4 & \textbf{0.5534} & \textbf{0.5534} & \textbf{0.5534} & \textbf{0.5534} \\
\hline
5 & \textbf{0.5204} & 0.5280 & 0.5280 & - \\
\hline
10 & \textbf{0.4102} & 0.4290 & 0.4290 & - \\
\hline
20 & \textbf{0.3486} & 0.3539 & 0.3543 & - \\
\hline
50 & \textbf{0.2261} & 0.2284 & 0.2282 & - \\
\hline
100 & \textbf{0.1501} & 0.1510 & 0.1507 & - \\
\hline
\end{tabularx}
\caption{The Mean Absolute Error of Radius Estimation (R=1000 mm) of the Sphere
Angle Ranges: Polar [0 90°], Azimuthal [0 360°]; $\sigma_s = 1 \, \text{mm}; \, \sigma_m = 1 \, \text{mm}$
}
\end{table}

\begin{table}[h]
\centering
\begin{tabularx}{0.48\textwidth}{|>{\centering\arraybackslash}X|>{\centering\arraybackslash}X|>{\centering\arraybackslash}X|>{\centering\arraybackslash}X|>{\centering\arraybackslash}X|}
\hline
\multirow{2}{=}{\centering Number\\of points} & \multicolumn{4}{c|}{Method} \\
\cline{2-5}
 & Our method formula (30), mm & Eberly method, mm & Sumith method, mm & Carnot formula, mm \\
\hline
4 & \textbf{0.4470} & \textbf{0.4470} & \textbf{0.4470} & \textbf{0.4470} \\
\hline
5 & \textbf{0.4416} & 0.4462  & 0.4462 & - \\
\hline
10 & \textbf{0.2999} & 0.3218 & 0.3218 & - \\
\hline
20 & \textbf{0.2190} & 0.2216 & 0.2216 & - \\
\hline
50 & \textbf{0.1329} & 0.1346 & 0.1347 & - \\
\hline
100 & \textbf{0.0921} & 0.0922 & 0.0922 & - \\
\hline
\end{tabularx}
\caption{The Mean Absolute Error of Radius Estimation (R=1000 mm) of the Sphere
Angle Ranges: Polar [0 120°], Azimuthal [0 360°]; $\sigma_s = 1 \, \text{mm}; \, \sigma_m = 1 \, \text{mm}$}
\end{table}

\begin{table}[h]
\centering
\begin{tabularx}{0.48\textwidth}{|>{\centering\arraybackslash}X|>{\centering\arraybackslash}X|>{\centering\arraybackslash}X|>{\centering\arraybackslash}X|>{\centering\arraybackslash}X|}
\hline
\multirow{2}{=}{\centering Number\\of points} & \multicolumn{4}{c|}{Method} \\
\cline{2-5}
 & Our method formula (30), mm & Eberly method, mm & Sumith method, mm & Carnot formula, mm \\
\hline
4 & 1.7031 & \textbf{0.5363} & 651.806 & 861.659 \\
\hline
5 & 1.6516 & \textbf{0.5853}  & 187.649 & - \\
\hline
10 & \textbf{0.3245} & 0.3880 & 0.3426 & - \\
\hline
20 & \textbf{0.2001} & 0.2094 & 0.2094 & - \\
\hline
50 & \textbf{0.1205} & 0.1220 & 0.1220 & - \\
\hline
100 & \textbf{0.0875} & 0.0886 & 0.0886 & - \\
\hline
\end{tabularx}
\caption{The Mean Absolute Error of Radius Estimation (R=1000 mm) of the Sphere
Angle Ranges: Polar [0 180°], Azimuthal [0 360°]; $\sigma_s = 1 \, \text{mm}; \, \sigma_m = 1 \, \text{mm}$}
\end{table}

In all cases except two, our method showed the best result. Let's consider these two cases. In both cases, the upper limit of the polar angle is 180°. In the first case, four points are considered, in the second case, five. The reason for the "failure" is that with a polar angle of 180° and a small number of points $N=4$ or 5, the first group of 4 points, which should form a tetrahedron, degenerates into a straight line with one point at the north pole and three points at the south pole. To ensure that our method shows the best result in these two cases as well, it is sufficient to move one of the points from the south pole to the north pole. Then, two pairs from the first group of points will form a degenerate regular tetrahedron. As shown earlier, such a tetrahedron is optimal for our method. To verify this statement, simply uncomment the operators 59, 60, and 61 in the \texttt{SphereRadiusFromDistance.m} function and repeat the calculation.

Let's consider another application of our method. Instead of measuring the linear distances between pairs of points, let’s assume we have access to measurements of pairwise arcs between points. This scenario is likely when the sphere has a large radius. Then, measuring all the linear distances is not possible due to the lack of measuring tools with the required parameters. In this case, a tape measure can be used to measure the arcs between pairs of points.

We apply our method to estimate the sphere's radius. We express half the square of the measured distance $C_{ij}$ between pairs of points through the arc length $L_{ij}$ between them:
\[
\frac{1}{2} C_{ij}^2 = 2R^2 \left( \sin \frac{L_{ij}}{2R} \right)^2. \tag{53}
\]
Substituting (53) into (30), we obtain an equation whose root is the radius $R$:
\[
\bar{b}^T A \bar{b} = 2, \tag{54}
\]
where the elements of the matrix $A_{ij} = \left( \sin \frac{L_{ij}}{2R} \right)^2$.

Formula (54) is analogous to formula (30). In (54), measured pairwise arcs are used to estimate the sphere's radius, whereas in (30), pairwise distances are used.

We can illustrate the radius estimation of the sphere using equation (54) by calculating the Earth's radius based on flight distances between cities. Flight trajectories are approximated by arcs of the Earth's surface. The flight distance data is taken from the website \url{https://www.travelmath.com/} and shown in Table VII.

\begin{table}[h]
\centering
\begin{tabularx}{0.48\textwidth}{|>{\centering\arraybackslash}X|>{\centering\arraybackslash}X|>{\centering\arraybackslash}X|>{\centering\arraybackslash}X|>{\centering\arraybackslash}X|>{\centering\arraybackslash}X|>{\centering\arraybackslash}X|>{\centering\arraybackslash}X|}
\hline
 & SEA & MOW & TYO & DEL & MVD & ANC & SYD \\
\hline
SEA & 0 & 8397 & 7712 & 11332 & 11258 & 3639 & 12454 \\
\hline
MOW &  & 0 & 7501 & 4341 & 13348 & 7017 & 14485 \\
\hline
TYO &  &  & 0 & 5850 & 18576 & 5572 & 7792 \\
\hline
DEL &  &  &  & 0 & 15598 & 9175 & 10419 \\
\hline
MVD &  &  &  &  & 0 & 13522 & 11882 \\
\hline
ANC &  &  &  &  &  & 0 & 11802 \\
\hline
SYD &  &  &  &  &  &  & 0 \\
\hline
\end{tabularx}
\caption{Flight Distances Between Cities (km) using IATA airport codes for cities: SEA (Seattle), MOW (Moscow), TYO (Tokyo), DEL (Delhi), MVD (Montevideo), ANC (Anchorage), SYD (Sydney)}
\end{table}

The program implementing the example is located in the file \texttt{Earth\_radius.m}. In the calculations, we assumed $\sigma_s = (10/3)$ $km$; $\sigma_m = (10/3)$ $km$. The results of the Earth's radius calculation are presented in Table VIII.

\begin{table}[h]
\centering
\begin{tabularx}{0.48\textwidth}{|>{\centering\arraybackslash}X|>{\centering\arraybackslash}X|>{\centering\arraybackslash}X|}
\hline
Number of cities & Earth radius (km) & Earth radius standard deviation (km) \\
\hline
4 & 6391.95 & 10.8985 \\
\hline
5 & 6375.08 & 2.63666 \\
\hline
6 & 6376.24 & 2.59656 \\
\hline
7 & 6374.57 & 2.00055 \\
\hline
\end{tabularx}
\caption{Estimation of the Earth's Radius Based on Flight Distances Between Cities}
\end{table}

According to Wikipedia, it is generally considered that the Earth has the shape of a sphere with an average radius of 6371.3 km. The calculations presented in Table VIII show that our method for estimating the Earth's radius is in good agreement with this value.

\section{Conclusion}

We have developed a method for determining the radius of a spherical surface with shape
deviations. The method is based on measuring pairwise distances between points fixed on the
measured surface. We obtained new analytical expressions for the radius of the sphere through
the matrix of pairwise distances. We determined the standard deviation of the radius estimate
caused by measurement errors and deviations of the sphere from the shape. We found optimal
configurations of points on the sphere that ensure the minimum value of the standard deviation
of the radius estimate.

\bibliographystyle{IEEEtran}
\bibliography{SphereRadius}

\end{document}